\documentstyle[11pt,aps,preprint,epsf,floats]{revtex}

\def\acal{{\cal A}}
\def\lcal{{\cal L}}
\def\ocal{{\cal O}}
\def\vcal{{\cal V}}
\def\gev{\hbox{GeV}}
\def\tev{\hbox{TeV}}
\def\up#1{^{\left( #1 \right) }}
\def\gesim{\,{\raise-3pt\hbox{$\sim$}}\!\!\!\!\!{\raise2pt\hbox{$>$}}\,}
\def\aa{{\bf a}}

\def\kk{{\bf k}}
\def\lll{{\bf l}}

\def\nn{{\bf n}}

\def\pp{{\bf p}}
\def\qq{{\bf q}}
\def\rr{{\bf r}}

\def\xx{{\bf x}}

\def\AA{{\bf A}}

\def\ui{U(1)}
\def\su#1{{SU(#1)}}
\def\tr{ \hbox{tr}}
\def\half{{1\over2}}
\def\inv#1{{1\over#1}}
\def\sqr#1#2{{\vcenter{\hrule height.#2pt
      \hbox{\vrule width.#2pt height#1pt \kern#1pt
         \vrule width.#2pt}
      \hrule height.#2pt}}}
\def\square{\mathchoice\sqr56\sqr56\sqr{2.1}3\sqr{1.5}3}

\def\qwe{f}
\def\leff{{\lcal}_{\rm eff}}

\def\bj{b-\hbox{jet}}
\def\cj{c-\hbox{jet}}
\def\eb{\epsilon_b}
\def\tc{t_c}
\def\tj{t_j}
\def\el{effective Lagrangian}
\def\sm{standard model~}

\begin{document}

\title{A short course in \el s.~\thanks{Lectures delivered at the
{\sl VII Mexican Workshop on Particles and Fields},                                                          
Merida, Yucatan, Mexico, 10-17 November, 1999.}}

\author{Jos\'e Wudka\thanks{{\tt jose.wudka@ucr.edu}}}
\address{Physics Department, UC Riverside \\ 
Riverside CA 92521-0413, USA}

\preprint{UCRHEP--T270}
\maketitle

\begin{abstract}
These lectures provide an introduction to effective 
theories concentrating on the basic ideas and providing some simple
applications
\end{abstract}

\section{Introduction.}

When studying a physical system it is often the case that there is not
enough information to provide a fundamental description of some of its
properties. In such cases one must parameterize the corresponding
effects by introducing new interactions with coefficients to be
determined phenomenologically. Experimental limits or measurement of
these parameters then (hopefully) provides the information needed to
provide a more satisfactory description. 

A standard procedure for doing this is to first determine the dynamical 
degrees of freedom involved and the symmetries obeyed, and then
 construct the most general Lagrangian, the 
{\em \el} for these degrees of freedom which respects
the required symmetries. The method is straightforward, quite general and,
most importantly, {\em it works!}

In following this approach one must be wary
of several facts. Fist it is clear that the relevant degrees of freedom
can change with scale (e.g. mesons are a good description of low-energy
QCD, but at higher energies one should use quarks and gluons); in
addition, physics at different scales may respect different symmetries
(e.g. mass conservation is violated at sufficiently high energies). It
follows that the \el\ formalism is in general applicable only for a
limited range of scales. It is often the case (but no always!) that
there is a scale $ \Lambda $ so that the results obtained using an \el\
are invalid for energies above $ \Lambda $.

The formalism has two potentially serious drawbacks. First,
\el\ has an infinite number of terms suggesting a lack
of predictability. Second, even though the model has an UV cutoff $
\Lambda $ and will not suffer from actual divergences, simple
calculations show that is is a possible for this type of theories to
generating radiative corrections
that grow with $ \Lambda $, becoming increasingly important
for higher and higher order graphs. Either of these problems can render
this approach useless. It is also necessary verify that the
model is unitary. 

I will discuss below how these problems are solved, an provide several
applications of the formalism. The aim is to give a flair of the
versatility of the approach, not to provide an exhaustive review of all
known applications.

 \section{Familiar examples}

\subsection{Euler-Heisenberg \el}

This Lagrangian summarizes QED at low energies (below the electron
mass)~\cite{iz}. At these energies 
only photons
appear in real processes and the \el~will be then constructed using the
photon field $A_\mu$, and will satisfy a $\ui$ gauge and Lorenz
invariances. Thus it can be constructed in terms of the field strength 
$F_{\mu \nu} $ or the loop variables
 $ \acal(\Gamma )= \oint_\Gamma A\cdot dx $. The latter are
non-local, so that a local description would involve only $F$, 
namely~\footnote{ There is no 
$F\tilde F $ terms since it is a total derivative.}
\begin{eqnarray}
\leff&=&\leff(F) \cr
&=& a F^2 + b F^4 + c (F\tilde F)^2 + d F^2 (F\tilde F) \cdots
\end{eqnarray}
One can arbitrarily normalize the fields and so choose
$ a=-1/4$. The constants
$b,~c$ and $d$ have units of mass$^{-2}$.

\begin{figure}
\vbox to 1.4in{\epsfxsize=4 in\epsfbox[-200 -600 412 192]{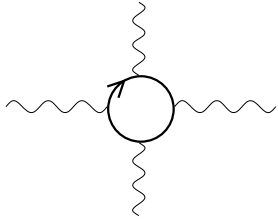}}
\label{f1}
\caption{Graph generating the leading terms in the Euler-Heisenberg \el}
\end{figure}

Note that the term $ \propto d$ violates CP. Though 
we know QED respects C and P, it is possible for other interactions
to violate these symmetries, there is nothing in the
discussion above that disallows such terms and, in fact, 
weak effects will generate them. For this system
we are in a privileged position for we know the underlying
physics, and so we can calculate $b,~c,~d,\ldots$. The
leading effects come form QED which yields
$ b,c\sim 1/(4 \pi m_e)^2 $ at 1 loop~\cite{iz}. The parameters
$b$ and $c$ summarize all the leading {\em virtual} electron effects.
(see Fig.~\ref{f1}). Forgetting about this underlying structure we
could have simply {\em defined} a scale $M$ and taken $b,~c \sim 1/M^2 $
(so that $ M = 4 \pi m_e $), and while this is perfectly viable,
$M$ is not relevant phenomenologically speaking as it does not
corresponds of a physical scale. In order to
extract information about the physics underlying the effective
Lagrangian from a measurement of $b$ and $c$ we must be able to
at least estimate the relation between these constants and the
underlying scales.

In addition we also know that $ d \sim \xi/(4 \pi v)$ with $v\sim 246
\gev$ and $\xi$ is a very small constant proportional to the Jarlskog
determinant~\cite{jarlskog}. The \el~can hold terms with radically
different scales and limits on some constants cannot, in general,
translate to others. In this case the terms are characterized by
different CP transformation properties, and it is often the case that
such global symmetries are useful in differentiating terms in 
the \el. The point being that a term violating
a given global symmetry at scale $\Lambda$ will generate
all terms in the \el~with the same symmetry properties through radiative corrections.
The caveat in the argument being that the underlying theory might have some
additional symmetries not apparent at low energies which might further
segregate interactions and so provide different scales for
operators with the same properties under all low energy symmetries.

When calculating with the \el~the effects produced by the
new terms proportional to
$b,c$ are suppressed by a factor $ \sim (E/4 \pi m_e)^2 $, where
$E$ is the typical energy on the process and $ E \ll m_e$. Thus the
effects of these terms are tiny, yet they are noticeable because
they generate a {\em new} effect: $ \gamma-\gamma$ scattering.

\subsection{(Standard) Superconductivity}

This is a brief summary of the very nice treatment provided by
Polchinski~\cite{polchinski}. The system under consideration
has the electron field $\psi$ as its only
dynamical variable (the phonons are assumed to have been 
integrated out, generating a series of electron self-interactions),
it respects $\ui$ electromagnetic gauge invariance,
as well as Galilean invariance and Fermion number conservation.

Assuming a local description, the first few terms in the
\el~expansion are (neglecting those
containing photons for simplicity)
\begin{equation}
 \leff = \int_k \psi^*_\kk \left[ i \partial_t - e_\kk + \mu \right] 
\psi_\kk + \int \psi^*_\kk \psi_\lll \psi_\qq \psi^*_\pp 
\delta(\kk-\lll-\qq+\pp) V_{\kk\lll\qq}+ \cdots
\end{equation}

In this equation the relation $e_\kk=\mu$ determines the Fermi surface, while
$ V\sim { (\hbox{electron-photon~coupling})^2 \over (\hbox{phonon~mass})^2}$ summarizes
the virtual phonon effects. In order to determine the importance of the
various terms we need the dimensions of the field $ \psi $. A vector \kk\ lies
on the Fermi Surface (FS) if $ e_\kk = \mu $, if \pp\ is near 
the FS one can write $ \pp = \kk + \ell \hat\nn $ (with $ e_\kk = \mu $). 
Scaling towards the FS implies $ \ell \to s \ell $ with $ s \to 0 $. Then
assuming $ \psi \to s^d \psi $ the quadratic terms in the action will be
scale invariant provided $d=-1/2$. The quartic terms in the
action then scales as $s$ and becomes negligible near the FS {\em except} 
when the pairing condition
$ \qq +\lll=0 $ is obeyed. In this case the quartic term scales as $
s^0$ and cannot be ignored. In fact this term
determines the most interesting behavior of the system at low temperatures
(see~\cite{polchinski} for full details).

\subsection{Electroweak interactions}

Again I will follow the general recipe. I will concentrate only on the
(low energy) interactions involving lepton fields, which are then the 
degrees of freedom. Since I assume the energy to be well below the Fermi
scale,
the only relevant symmetries are $\ui$ gauge and Lorenz invariances. In addition
there is the question whether the heavy physics will respect the discrete
symmetries $C$, $P$ or $CP$; using perfect hindsight I will retain terms
that violate these symmetries

Assuming a local description I have~\cite{iz}
\begin{equation}
\leff = \sum \bar\psi_i(i \not\!\!D -m_i)\psi_i +
\sum \qwe_{ijkl} \left(\bar\psi_i \Gamma^a \psi_j \right)
\left(\bar\psi_k \Gamma_a \psi_l\right) + \cdots 
\end{equation}
where the ellipsis indicate terms containing operators of
higher dimension, or those involving the electromagnetic field.
The matrices $ \Gamma$ are to be chosen among the 16
independent basis 
$\Gamma^a=\{1, \gamma_\mu, \sigma_{\mu\nu},\gamma_\mu\gamma_5,\gamma_5\}$

\begin{figure}[h]
\vbox to 1in{\epsfxsize=6.3 in\epsfbox[-200 -700 412 92]{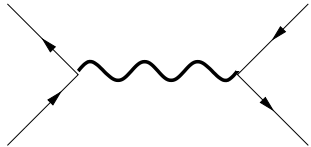}}
\label{f2}
\caption{Standard model processes generating four fermion interactions at low energies
(e.g.. Bhaba scattering)}
\end{figure}

The coefficients for the first two terms are be fixed by normalization
requirements. While a SM calculation gives
 $ \qwe \sim g^2/m_W^2=1/v^2 $ ($v\simeq246$GeV)
and is generated by tree-level
graphs (see Fig. \ref{f2}) because of this the scale
$1/\sqrt{\qwe}$ is, in fact, the scale of the heavy physics
and so the model is applicable at energies swell below $v$.
The four fermion interactions
summarize the leading virtual gauge boson effects.
The contributions of the four-fermion operators
to processes with typical energy $E$
are suppressed by a factor $ E^2/v^2$. These can be observed (or
bounded)
despite the $ E \ll v$ condition because they generate
{\em new} effects: $C$ and $P$ (and some of them chirality) violation.

\subsection{Strong interactions at low energies}

In this case we are interested in the description of the interactions among
the lightest hadrons, the meson multiplet. The most convenient parameterization
of these degrees of freedom is in terms of a unitary field\cite{georgi} 
$U$ such that
 $ U=\exp(\lambda_a\pi^a/F)$ where $\pi^a$ denote the eight meson fields, $ \lambda^a $ 
the Gell-Mann matrices and $F$ is a constant (related to the pion decay 
constant).
The symmetries obeyed by the system are chiral $\su3_L\times\su3_R$, Lorenz invariance, $C$ 
and $P$.

With these constraints the \el~takes the form
\begin{equation}
\leff = a\; \tr \partial U^\dagger \cdot \partial U + \left[ b\; \tr \partial_\mu U^\dagger
\partial_\nu U \partial^\mu U^\dagger \partial^\nu U + 
\cdots\right] + \cdots 
\end{equation}

I can set $a \sim F^2 $ by properly normalizing the
fields. In this case the leading term in the \el~will determine
all (leading) low-energy pion interactions in terms of the single constant $F$.
The effects form the higher-order terms have been measured and the data
requires $b\sim 1/(4\pi)^2$. This result is also predicted by the consistency
of this approach which requires that
radiative corrections to $a$, $b$, etc. should be at
most of the same size as their tree-level values.

 \section{ Basic ideas on the applicability of the formalism}

Being a model with intrinsic an cutoff there are no actual ultraviolet
divergences in most \el~computations. Still there are
interesting renormalizability issues that arise when doing effective
Lagrangian loop computations. 

Imagine doing a loop calculation including some vertices terms of (mass)
dimension higher than the dimension of space-time. These must have coefficients
with dimensions of mass to some negative power. The loop integrations
will produce in general terms growing with $ \Lambda $ the UV cutoff
which are  polynomials in the external momenta~\footnote{Since a
graph can be rendered convergent by taking sufficient number of derivatives
with respect to the external momenta.} 
and will preserve the symmetries
of the model~\cite{collins}. Hence these terms which may {\em grow}
with $ \Lambda $
correspond to vertices
appearing in the most general \el~ and can be absorbed in a
 renormalization of the corresponding coefficients. They have no
observable effects (though they can be used in naturality arguments~\cite{aew1}. 

Effective theories will also be unitary {\em provided} one stays within the limits
of their applicability. Should one exceed them new channels will open (corresponding
to the production of the heavy excitations) and unitarity violating effects
will occur. This is {\em not} produced by real unitarity violating
interactions, but due to our using the model beyond its range of
applicability (e.g. it the typical energy of the process
under consideration reaches of exceeds $ \Lambda $). One can, of course, {\em extend}
the model, but this necessarily introduces {\it ad-hoc} elements and will dilute
the generality gained using effective theories. 

For example consider $WWZ$ interactions with an \el~of the form
\begin{equation} 
\leff= \lambda(p,k)W_{\mu\nu}(k)W^{\nu\rho}(p)Z_\rho{}^\mu (-p-k)+
	\cdots;
\end{equation}
(where $V_{\alpha \beta}= \partial_\alpha V_\beta -\partial_\beta V_\alpha$)
One can then choose $ \lambda$ to insure unitarity is preserved (at least in
some processes), for example~\cite{bz}
\begin{equation}
\lambda(p,k)={\lambda_0\over(p\cdot k+\Lambda)^n}
\end{equation}
which, for $n$ sufficiently large insures that the cross section
for the reaction $e^+e^-\to Z \to WW$ is unitary, since it behaves as
$ s^{2-2n} $ for a CM energy$=s\gg\Lambda^2$. But 
the very same effective vertex also modifies other reactions such
as, for example $u \bar d \to W \to ZW$ where the cross section now has
a factor
$ ( s - \Lambda^2 )^{-2n}$ and will exhibit resonant behavior if $ s \sim
\Lambda^2 $. If one requires $ s \ll \Lambda^2 $ (as required by the consistency
of the formalism) there are neither unitarity violations nor
resonance effects. If, however, one uses the above Ansatz to extend the
range of applicability to $ s \sim \Lambda^2 $ and beyond then
very clear resonances should be observed in hadron colliders. Given that these
have not been observed one {\em must} use for $ \Lambda $ a value 
significantly larger than the average CM energy for the hard $W$ pair production
cross section. 

\section{Using effective Lagrangians}

Effective Lagrangians provide an efficient way
of summarizing some (perhaps very complex) interactions. The idea is simply
to include all the effective vertices produces by those excitations which
are not directly observed. 

For example given a real scalar field $ \phi $
and assume that all Fourier components above a scale $ \Lambda $ are not
directly observable ({\i.e.} the available energies lie all below $ \Lambda $), then 
the \el~is obtained by 
integrating over the variables observable at energies $ \ge \Lambda$; writing 
 $ \phi=\phi_0+\phi_1$, with
\begin{equation}
\phi_0(\kk):~|\kk|<\Lambda\qquad \phi_1(\kk):~\Lambda\le|\kk|<\Lambda_1
\end{equation}
then by definition
\begin{eqnarray} 
e^{i S_{\rm eff}} &=& \int [d\phi_1] e^{i S(\phi_0,\phi_1)}, \qquad
S_{\rm eff} = \int d^n x \leff
\end{eqnarray}
where $\leff$ is obtained by expanding $ S_{\rm eff}$
in powers of $\Lambda $ which gives an infinite tower of local operators.

Another common situation where \el s appear occurs when some heavy excitations are
integrated out. This can be illustrated by the following toy 
model~\footnote{I'm cheating in order to get a closed form for the effective action, 
a more realistic model should include a term $ \propto \phi_1^4 $}
\begin{equation}
 S = \int d^n x \left[ \bar \psi ( i \not\!\partial -m)\psi
+ \half (\partial \phi)^2 -\half \Lambda^2 \phi_1^2 + \qwe \phi\bar\psi\psi
\right] 
\end{equation}
where $ \phi$ is heavy. A
simple calculation gives
\begin{equation}
 S_{\rm eff} = \int d^n x \left[ \bar \psi ( i \not\!\partial -m)\psi
+ \half\qwe^2 \bar\psi\psi\inv{\square+\Lambda^2} \bar\psi\psi
\right] 
\end{equation}
and
\begin{equation}
 \leff = \bar \psi ( i \not\!\partial -m)\psi
+{\qwe^2\over 2 \Lambda^2}\sum_{l=1}^\infty \bar\psi\psi
\left(\square\over\Lambda^2\right)^n \bar\psi\psi
\end{equation}
Note that terms with large number of derivatives will be suppressed
by a large power of the small factor $ (E/\Lambda)$, if we are interested
in energies $ E \sim \Lambda $ the {\em whole} infinite set of vertices must
be included in order to reproduce the $\phi$ pole.

 \subsection{How to parameterize ignorance}

If one knows the theory we can, in principle, calculate $ \leff $ (or
do a full calculation). Yet there are many cases where the underlying theory
is not known. In these cases an effective theory if obtained
by writing {\em all} possible interactions among the light
excitations. The model then has an infinite number of terms each with an
unknown parameter, and these constants then parameterize {\em all} possible
underlying theories. The terms which dominate are those usually called
renormalizable (or, equivalently, marginal or relevant). The other terms
are called non-renormalizable, or irrelevant, since their effects become
smaller as the energy decreases

This recipe for writing effective theories must be supplemented with
some symmetry restrictions. The most important being that the
all the terms in the \el~must respect the local gauge
invariance of the low-energy physics (more technically, the one
respected by the renormalizable terms in the effective 
action)~\cite{iruv}. The reason is 
that the presence of a gauge variant term will generate {\em all} 
gauge variant interactions
thorough renormalization group evolution. 

\paragraph{Gauge invariantizing}

Using a simple argument it is possible to turn any theory into a gauge 
theory~\cite{bl} and so it appears that the
requirement of gauge invariance is empty. That this is not the case is
explained here. I first describe the trick which grafts gauge invariance
onto a theory and then discuss the implications.

Consider an arbitrary
theory with matter fields (spin 0 and 1/2) and vector
fields $V_\mu^n$, $n=1,\ldots N$. Then
\begin{itemize}
\item Choose a (gauge) group $G$ with $N$ generators $\{T^n\}$.
Define a covariant derivative
$D_\mu = \partial_\mu + V_\mu^n T^n$ and
{\em assume} that the $V_\mu^n$ are gauge fields.
\item Invent a unitary field $U$ transforming according to the fundamental 
representation of $G$ and construct the gauge invariant
composite fields
\begin{equation}
\vcal_\mu^n = -\tr T^n U^\dagger D_\mu U
\end{equation}
Taking $\tr T^n T^m = - \delta_{nm} $, it is easy to see that in the unitary gauge $U=1$, 
$\vcal_\mu^n = V_\mu^n $.
\end{itemize}

Thus if simply replace $V \to \vcal $ in the original
theory we get a gauge theory. Does this mean that gauge invariance irrelevant
since it can be added at will? In my opinion this is not the case. 

In the above process
{\em all matter fields are assumed gauge singlets} (none are minimally
coupled to the gauge fields).In the case of the \sm, for example, the
universal coupling of fermions to the gauge bosons would be accidental
in this approach. In order to recover the full predictive power
commonly associated with gauge theories, the matter fields must transform
non-trivially under $G$ which can be done only if there are strong correlations
among some of the couplings. It is {\em not} trivial to say that the \sm\ group is 
 $\su3\times\su2\times\ui$ with left-handed quarks transforming as
$(\boldmath{3,2,1/6})$, left-handed leptons as $(\boldmath{1,2,-1/2})$,
etc., as opposed to a $\ui^{12}$ with all fermions transforming as
singlets~\cite{jw}.

\subsection{How to estimate ignorance}

A problem which I have not addressed so far is the fact that
effective theories have an infinite number of coefficients,
with the (possible) problem or requiring an infinite number of data points
in order to make any predictions. On the other hand, for example, 
if this is the case why is it
that the Fermi theory of the weak interactions is so successful?

The answer to this question lies in the fact that not all coefficients are created 
equal, there is a {\em hierarchy}~\cite{georgi,jw}. As a result, given any desired
level of accuracy, only a finite number of terms need to be included.
Moreover, even though the effective
Lagrangian coefficients cannot be calculated without knowing the
underlying theory, they can still be {\em bounded} using but a minimal
set of assumptions about the heavy interactions. It is then also
possible to estimate the errors in neglecting all but the finite number
of terms used.

As an example consider the \sm\ at low energies and calculate two processes: Bhaba 
cross section and the anomalous magnetic moment of the electron.
For Bhaba scattering there is a contribution due the $Z$-boson exchange
(see Fig. \ref{f2})
\begin{equation}
 e^+ e^- \to Z \to e^+e^- \quad \hbox{generates} \quad \ocal = \inv{2 m_Z^2}
\left(\bar e \Gamma \gamma^\mu e\right)
\left(\bar e \Gamma \gamma_\mu e\right)
\end{equation}
where $\Gamma = g_V + g_A \gamma_5 $. The coefficient of the
effective operator $ \ocal $ is then
$ \sim (\hbox{coupling}/\hbox{physical~mass})^2\sim 1/ v^2 $

\begin{figure}
\vbox to 3in{\epsfxsize8 in\epsfbox[-100 -560 512 232]{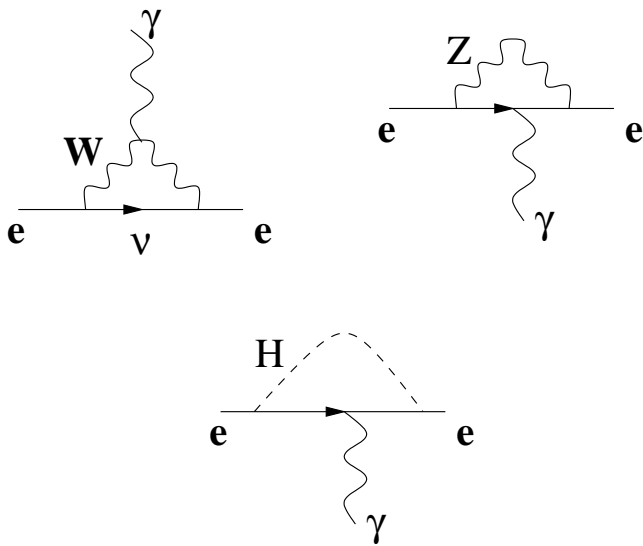}}
\label{f5}
\caption{Weak contributions to the electron anomalous magnetic moment}
\end{figure}

The electron anomalous magnetic moment receives
contributions from virtual $W$, $Z$ and $H$ exchanges
(see Fig. \ref{f5}). The corresponding low-energy
operator is
\begin{equation}
 \ocal = \bar e \sigma_{\mu\nu} e F^{\mu\nu}
\end{equation}
In this case the coefficient $ \sim \{ \hbox{coupling}/[4 \pi
(\hbox{physical~mass})]\}^2 \sim 1/(4 \pi v)^2$~\footnote{In addition
the coefficient is suppressed by a factor of $m_e$ since it violates
chirality.}.

The point of this exercise is to illustrate the fact that,
for weakly coupled theories, loop-generated operators
have smaller coefficients than operators generated at tree level.
Leading effects are produced by operators 
which are generated at tree level.

\subsection{Coefficient estimates}

In this section I will provide arguments which can be
used to estimate (or, at least bound) the coefficients in the
\el. These are order of magnitude
calculations and might be off by a factor of a few; it 
is worth noting that no single
calculation has provided a significant deviation from
these results.

The estimate calculations should be done separately for
weakly and strongly interacting theories. I will
characterize the first as those where radiative corrections 
are smaller than the tree-level contributions. Strongly
interacting theories will have radiative corrections of
the same size at any order~\footnote{Should the radiative
corrections increase with the order of the calculation, it is
likely that the dynamic variables being used are not appropriate
for the regime where the calculation is being done.}

\subsubsection{Weakly interacting theories}

In this case leading terms in the \el~are those
which can be generated at tree level by the heavy physics. Thus the
dominating effects are produced by operators which have the lowest
dimension (leading to the smallest suppression from inverse powers of
$ \Lambda $) and which are tree-level generated (TLG) operators can 
be determined~\cite{aew2}. 

When the heavy physics is described by a gauge theory it
is possible to obtained all TLG operators~\cite{aew2}. 
The corresponding vertices fall into 3 categories, 
symbolically
\begin{itemize}
\item vertices with $4$ fermions.
\item vertices with $2$ fermions and $k$ bosons; $k=2,3$
\item vertices with $n$ bosons; $n=4,6$.
\end{itemize}
 A particular
theory may not generate one or more of these vertices, the only
claim is that there is {\em a} gauge theory which does.

In the case of the \sm\ with lepton number conservation the leading
operators have dimension 6~\cite{bw,aew2}.
Subleading operators are either dimension 8 and their contributions
are suppressed by an additional factor $(E/\Lambda)^2 $ in processes with
typical energy $E$. Other subleading contributions are suppressed by
a loop factor $\sim 1/(4\pi)^2$. Note that it is possible to have situations
where the only two effects are produced by either dimension 8 TLG
operators or loop generated dimension 6 operators. In this case
the former dominates only when $\Lambda > 4 \pi E $.

\paragraph{Triple gauge bosons}

The terms in the electroweak \el~which describe the
interaction of the $W$ and $Z$ bosons generated by
some heavy physics underlying the \sm has received considerable
attention recently~\cite{ellison}. In terms of the $\su2$ and $\ui$ gauge fields
$W$ and $B$ and the scalar doublet $ \phi$ these interactions are
\begin{eqnarray}
\leff &=& {1\over \Lambda^2} \left( \alpha_W \ocal_W + 
\alpha_{BW} \ocal_{BW} \right) \cr
\ocal_W &=& \epsilon_{IJK} W^I_{\mu\nu} W^J{}^\nu{}_\lambda W^K{}^{\lambda\mu} \cr
\ocal_{WB} &=& \phi^\dagger \tau^I \phi W^I_{\mu\nu} B^{\mu\nu}
\end{eqnarray}

The above arguments inly that there is no TLG operator 
containing three gauge bosons.
This means that all effective contributions to the $WWZ$ and 
$WW\gamma$ interactions are loop generated, so their
coefficients {\em necessarily} take the form $ \prod
(\hbox{coupling~constants})/(16 \pi^2)$. Thus the parameters
$\kappa$ and $ \lambda $ commonly used to parameterize these interactions
are of order $ 5 \times 10^{-3} $. {\em Experiments
providing limits significantly above this value provide
\underline{no} information about the heavy physics}.

\subsubsection{Strongly interacting theories}

I will imagine a theory containing scalars and fermions
which interact strongly. Gauge couplings are assumed to be small
and will be ignored. This calculation is useful for
low energy chiral theories but not for low energy QCD~\cite{gl,gm,georgi}.

A generic effective operator in this type of theories takes
the form
\begin{equation}
 \ocal_{abc} \sim \lambda \Lambda^4
\left({ \phi\over \Lambda_\phi}\right)^a 
\left({ \psi\over \Lambda_\psi}^{3/2}\right)^b 
\left({ \partial\over \Lambda}\right)^c
\end{equation}
Then the condition that these dynamic variables appropriately
describe the physics below $\Lambda $ implies that
radiative corrections to the couplings are at most
as large as the tree-level values, namely
$ \delta_{\rm rad} \lambda \le \lambda $. A straightforward
estimate (including a factor of $1/(16\pi^2)$ for each loop)
shows that this condition is satisfied only if
\begin{equation}
 \Lambda_\psi = \inv{(4\pi)^{2/3}} \Lambda, \quad 
\Lambda_\phi = \inv{4 \pi}\Lambda,
\quad \lambda = \inv{16 \pi^2} 
\end{equation}

In terms of $U\sim \exp(\phi/\Lambda_\phi)$, the operators take the form
\begin{equation}
\ocal_{abc} = \inv{(4\pi)^{2-b}} \Lambda^{4-c-3b/2} \partial^c U^{a'} \psi^b
\end{equation}
In particular the coefficient of the two derivative operators
$ \tr\partial U^\dagger \partial U  $ is $ \propto \Lambda_\phi^2$.

For the case where $ \phi$ represents the interpolating field for
the lightest mesons PCAC implies
$\Lambda_\phi = f_\pi $~\cite{gl,georgi}. Then
\begin{equation}
\psi^4 \propto \inv{f_\pi^2}\qquad
\partial^4 U^4 \propto \inv{16\pi^2}\qquad
\psi^2 \partial^2 U^2 \propto \inv{4 \pi f_\pi}
\end{equation}
(note that these are upper bounds). The extensive data
on low energy meson reactions can be used to gauge the
validity of these predictions, they are indeed satisfied. In 
particular the $ (\partial U)^4$ terms have coefficients $ \sim
1/(16\pi^2) $.

For the case of the \sm the field $U$ can be used
to provide masses for the $W$ and $Z$ bosons without a physical
Higgs being present (the price is that the model breaks down at
energies $ \sim 4 \pi v =3$TeV). In this case the gauge
fields are introduced minimally and it is the term
$(D U)^2 $ gives a mass to the $W$ and $Z$ which fixes
$\Lambda_\phi = v =246\gev $ whence $\Lambda = 3\tev $;
as before, the model makes no sense beyond this 
scale~\footnote{Tough it is conceivable that a full
non-perturbative calculation would show that the theory cures
itself and can be extended beyond this scale, there is no indication
that this miracle occurs.} In addition, when the gauge fields are
reintroduced, the terms with 4 derivatives
will generate triple-vector boson couplings, again leading to the
estimates $ \lambda, \kappa \sim 5 \times 10^{-3} $~\cite{jw}.

\subsection{Radiative corrections}

Despite the presence of higher-dimensional operators radiative
corrections can be calculated in the usual way. As an example
imagine calculating the corrections to the cross section for
the reaction $ e^+ e^- \to e^+ e^- $ using the \sm
with the addition of a 4-fermion interaction
\begin{equation}
\leff = \leff^{\rm SM} + {f \over \Lambda^2 } \left( \bar \psi
\gamma^\mu \psi \right) \left( \bar \psi \gamma_\mu \psi \right)
+ \cdots 
\end{equation}
where $\psi$ denotes the electron field. 

\begin{figure}
\vbox to 3in{\epsfxsize 6 in\epsfbox[0 -450 612 342]{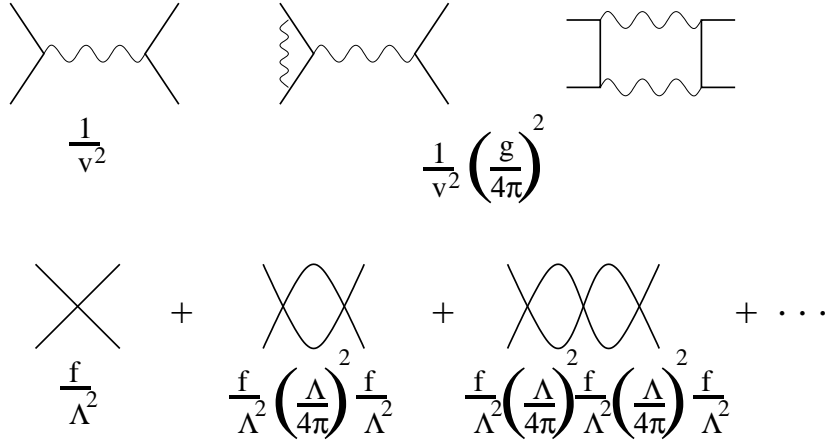}}
\label{f10}
\caption{Radiative corrections to Bhaba scattering in the presence
of a 4-fermion interaction}
\end{figure}

The calculation is illustrated in Fig. \ref{f10} where the loops
involving the 4-fermion operator are cut-off at a scale $ \Lambda $. The 
SM and new physics (NP) contributions are, symbolically,
\begin{eqnarray}
\hbox{SM:}&&{1\over v^2} \left[1+ {g^2 \over 16 \pi^2} + \cdots \right] \cr
\hbox{NP:}&&{f\over \Lambda^2}\left[1 + {f \over 16\pi^2} + \cdots
\right]
\end{eqnarray}
Note that this consistent behavior (that the new physics effects
disappear as $ \Lambda \to \infty $) results form having the physical
scale of new physics $ \Lambda$ in the coefficient of the operator. Had
we used $ \qwe'/v^2 $ instead of $ \qwe/\Lambda^2 $ the new physics
effects would appear to be enormous, and growing with each new loop. It
is not that the use of $ \qwe'/v^2 $ is wrong, it is only that it is
misleading to believe $ \qwe' $ can be of order one; it {\em must} be
suppressed by the small factor $ (v/\lambda)^2 $.

Using these results we see that this reaction is sensitive
to $ \Lambda $ provided $ f (v/\Lambda)^2 >$sensitivity. If the
sensitivity is, say 1\%\ this corresponds to $ \Lambda/\sqrt{f} >
2.5$TeV~\cite{grzadkowski}.

 This perturbative calculation is manageable provided
$ f < 16 \pi^2 $, otherwise the underlying physics is strongly
coupled. It is still possible in that case to provide estimates
of the new physics contributions, though these are less reliable,
these estimates imply that $1 + f/(4\pi)^2 + \cdots \sim 1 $ when
$ f \sim 16 \pi^2 $.

\section{Applications to electroweak physics}

With the above results one can determine, for any given process,
the leading contributions (as parameterized by the various effective operator
coefficients). Using then the coefficient estimates one can provide 
the expected magnitude of the new physics effects with only $ \Lambda $ 
as an unknown parameter, and so estimate the sensitivity to the scale of new physics.

It is important to note that this is sometimes a rather involved calculation as
all contributing operators must be included. For example, in order to determine the
heavy physics effects on the oblique parameters one must calculate not only these
affecting the vector boson polarization tensors, but also this which modify the Fermi
constant, the fine structure constant, etc. as these quantities are used when
extracting $S$, $T$ and $U$ from the data~\cite{sanchez}.

 \subsection{ Effective lagrangian}

In the following I will assume that the underlying physics is weakly coupled 
and derive the
leadingoperators that can be expected form the existence of heavy excitations
at scale $ \Lambda $.

The complete list of dimension 6 operators was cataloged a long
time ago for the case where the low energy spectrum includes
a single scalar doublet~\cite{bw}~\footnote{More complicated scalar sectors have
also been studied~\cite{perez}, though not exhaustively.}. It is then straightforward
to determine the subset of operators which can be TLG, they
are~\cite{aew2}
\begin{itemize}
\item{Fermions:}~$\left( \bar\psi_i \Gamma^a \psi_j \right)\left( \bar\psi_k \Gamma^a \psi_l \right)$
\item{Scalars:}~$|\phi|^6,~ (\partial |\phi|^2)^2$
\item{Scalars~and~fermions:}~$|\phi|^2 \times \hbox{Yukawa~term}$
\item{Scalars~and~vectors:}~$|\phi|^2 |D\phi|^2, ~ |\phi^\dagger D\phi|^2$
\item{Fermions,~ scalars~and~vectors:}~ 
$\left(\phi^\dagger T^n D^\mu \phi \right)\left(\bar\psi_i T^n \gamma_\mu \psi_j \right)$
\end{itemize}
where $T$ denotes a group generator and $ \Gamma$ a product of 
a group generator and a gamma matrix.

Observables affected by the operators in this list provide the highest
sensitivity to new physics effects provided that the \sm\ effects are
themselves small (or that the experimental sensitivity is large enough
to observe small deviations). I will illustrate this with two (incomplete)
examples

 \subsection{ b-parity}

This is a proposed method for probing new flavor physics~\cite{barshalom}. 
Its virtue lies in 
the fact that it is very simple  and sensitive (though it does not provide
the highest sensitivity for all observables). The basic idea is based on the
observation that the \sm acquires an additional global $\ui_b$ symmetry
in the limit $V_{ub}=V_{cb}=V_{td}=V_{ts}=0$ (given the experimental 
values
$0.002 < |V_{ub}| < 0.005$, 
$0.036 < |V_{cb}| < 0.046$,
$0.004 < |V_{td}| < 0.014$,
$0.034 < |V_{ts}| < 0.046$,
deviations form exact $\ui_b$ invariance will be small). Then for any
\sm interaction a reaction to the type
\begin{equation}
 n_i~\bj +X \to n_f~\bj+Y
\end{equation}
will obey
\begin{equation}
(-1)^{n_i} = (-1)^{n_f}
\end{equation}
to very high accuracy. The number $(-1)^{{\rm\#~of~}b~{\rm jets}}$ 
defines the {\bf b-parity} of a state (it being understood that the
top quarks have decayed).

The \sm is then b-parity even, and the idea is to consider a lepton
collider~\footnote{In hadron colliders there are sea-$b$ quarks which foul-up the
argument} and simply count the number of $b$ jets in the final state;
new physics effects
will show up as events with odd number of $b$ jets. 

The \sm\ produces no measurable
irreducible background, yet there are significant {\em reducible} backgrounds which
reduced the sensitivity to $ \Lambda $. To estimate these effects I define
\begin{itemize}
\item $\eb=\bj$ tagging efficiency
\item $\tc=\cj$ {\bf mis}tagging efficiency (probability of mistaking a $\cj$ jet for a $\bj$
\item $\tj=$light-jet {\bf mis}tagging efficiency (probability of mistaking a light-jet for a $\bj$
\end{itemize}
so that the {\em measured} cross section with $k$-\bj s is
\begin{equation} \!\!\!\!\!\!\!\!\!\!\!\!\!
\bar\sigma_k = \sum_{u+v+w=k} 
\left[ {n \choose u} \eb^u (1-\eb)^{n-u} \right] 
\left[ {m \choose v} \tc^v (1-\tc)^{m-v} \right] 
\left[ {\ell \choose w} \tj^w (1-tj)^{\ell-w} \right] 
\sigma_{nm\ell} 
\end{equation}
where $ \sigma_{nm\ell}$ denotes the cross section for the final state
with $n$ \bj s, $m$ \cj s, and $\ell$ light jets. Note that
$ \left[ {n \choose u} \eb^u (1-\eb)^{n-u} \right]$ is the probability of tagging
$u$ and missing $n-u$ \bj s out of the $n$ available.

As an example consider
\begin{equation}
\leff=\lcal_{\rm sm} +{\qwe_{ij}\over\Lambda^2} 
 \left(\bar \ell \gamma^\mu \ell\right)\left( \bar q_i \gamma_\mu q_j \right)
\end{equation}
where $i\not=j$ denote family indices. 
Taking $m_H=100\gev~|\qwe|=1~t_c=t_j=0$ the sensitivity
to $ \Lambda $ is summarized by the following table

\begin{center}
\begin{tabular}{||c|c|c|c|c||}
\hline
\multicolumn{5}{||c||}{Limits from $e^+e^- \to t \bar c + 
\bar t c +b \bar s + \bar b s \to 1b{\rm-jet} +X$}\\ \hline
$\sqrt{s}$ & 
$ L $&
$\epsilon_b=50\%$ &
{$\epsilon_b=60\%$}&
{$\epsilon_b=70\%$}\\ 
\hline
200~GeV&
2.5~$fb^{-1}$&
1.4~TeV&
1.5~TeV&
1.6~TeV
\\ \hline
500~GeV&
75~$fb^{-1}$&
5.0~TeV&
5.2~TeV&
5.5~TeV
\\ \hline
1000~GeV&
200~$fb^{-1}$&
9.5~TeV&
10.0~TeV&
10.7~TeV
\\
\hline
\end{tabular}
\end{center}

These results are promising yet they will be degraded
in a realistic calculation. First one must
include the effects of having $t_{c,j}\not=0$. In addition
there are complications in using inclusive reactions such as $ e^+ e^- \to b + X $
since the contributions form events with large number of jets can be
very hard to evaluate (aside from the calculational difficulties there are
additional complications when {\em defining} what a jet is). 
A more realistic approach 
is to restrict the calculation to a sample with a fixed number of jets ($2$
and $4$ are the simplest) and determine the sensitivity to $ \Lambda $ for
various choices of $ \eb$ and $\tj$ using this population only.

 \subsection{ CP violation}

Just as for b-parity the CP violating effects are small within the \sm 
and so precise measurements of CP violating observable might be very
sensitive to new physics effects. 

In order to study CP violations it is useful to first define what the
CP transformation {\em is}. In order to do this in general denote
the Cartan group generators by $H_i$ and the root generators by
$E_{\boldmath\alpha}$, then it is possible to find a basis where
{\em all} the group generators are real and, in addition, the $H_i$ are
diagonal~\cite{einhorn}. 
Define then CP transformation by
Transformations
\begin{eqnarray}
\psi & \to& C \psi^*~\hbox{(fermions)}\cr
\phi &\to& \phi^*~\hbox{(scalars)}\cr
A_\mu\up i &\to& -A_\mu\up i, ~(i:\hbox{~Cartan generator})\cr
A_\mu\up{\boldmath\alpha} &\to& - A_\mu\up{-\boldmath\alpha},~
({\boldmath\alpha}: \hbox{~root}) \nonumber
\end{eqnarray}
it is easy to see that the
field strengths and currents transform as $A_\mu$, while
$D\phi \to (D\phi)^* $. It then follows that in this basis
the whole gauge sector of {\em any} gauge
theory is CP conserving; CP violation can arise {\em only} in the
scalar potential and fermion-scalar interactions using this basis.

In order to apply this to electroweak physics I will need the list
of TLG operators of dimension
6 which violate CP, they are given by~\footnote{The
notation is the following: $\ell$ and $q$ denote the left-handed lepton and quark doublets; 
$u$, $d$ and $e$ denote the right handed quark and charged lepton fields. $lambda$ denote
the Gell Mann matrices, $\tau $ the Pauli matrices, and $ \epsilon = i \tau^2$.
$D$ represents the covariant derivatives and $\phi$ the scalar doublet.}
\begin{eqnarray}
&&\left(\bar\ell e\right)\left(\bar d q\right)-\hbox{h.c.}\qquad
\left(\bar q u\right) \varepsilon\left(\bar q d\right)-\hbox{h.c.}\qquad
\left(\bar q \lambda^A u\right) \varepsilon\left(\bar q \lambda^Ad\right)-\hbox{h.c.}\cr
&&\left(\bar \ell e\right) \varepsilon\left(\bar q u\right)-\hbox{h.c.}\qquad
\left(\bar \ell u\right) \varepsilon\left(\bar q e\right)-\hbox{h.c.}\qquad
|\phi|^2\left(\bar\ell e \phi-\hbox{h.c.}\right) \cr
&&|\phi|^2\left(\bar q u \tilde\phi-\hbox{h.c.}\right) \qquad
|\phi|^2\left(\bar q d\phi-\hbox{h.c.}\right) \qquad
|\phi|^2 \partial_\mu \left(\bar \ell \gamma^\mu \ell\right) \cr
&&|\phi|^2 \partial_\mu \left(\bar e \gamma^\mu e\right) \qquad\qquad
|\phi|^2 \partial_\mu \left(\bar q \gamma^\mu q\right) \qquad\quad
|\phi|^2 \partial_\mu \left(\bar u \gamma^\mu u\right) \cr
&&|\phi|^2 \partial_\mu \left(\bar d \gamma^\mu d\right) \cr
&&\ocal_1=\left(\phi^\dagger\tau^I\phi\right) D^{IJ}_\mu \left(\bar \ell \gamma^\mu \tau^J \ell\right) \cr
&&\ocal_2=\left(\phi^\dagger\tau^I\phi\right) D^{IJ}_\mu \left(\bar q \gamma^\mu \tau^J q\right) \cr
&&\ocal_3=\left(\phi^\dagger \varepsilon D_\mu\phi\right) \left(\bar u \gamma^\mu d \right) -\hbox{h.c} \nonumber
\end{eqnarray}

All operators except
$\ocal_{1,2,3}$ violate chirality and their coefficients are strongly bounded by their
contributions to the strong CP parameter $\theta$; in addition some chialiry violating
operators contribute to meson decays (which again provide strong bounds
for fermions in the first generation) and, finally, in natural theories some
contribute radiatively to fermion masses and will be then suppressed by
the smaller of the corresponding Yukawa couplings. For these reasons I will not
consider them further. Moreover, since I will be interested in limits that can be
obtained using current data, I will ignore operators whose only observable effects
involve Higgs particles. 

With these restrictions only $\ocal_{1,2,3}$ remain; their terms not involving scalars
are
\begin{eqnarray}
\ocal_1 &\to& -{i g v^2\over \sqrt{2}} \left(\bar \nu_L \not\!\!W^+ e_L -\hbox{h.c.}\right)\cr
\ocal_2 &\to& -{i g v^2\over \sqrt{2}} \left(\bar u_L \not\!\!W^+ d_L -\hbox{h.c.}\right)\cr
\ocal_3 &\to& -{i g v^2\over \sqrt{8}} \left(\bar u_R \not\!\!W^+ d_R -\hbox{h.c.}\right)\nonumber
\end{eqnarray}
The contributions from $\ocal_{1,2}$ can be absorbed in a renormalization of \sm coefficients
whence only $\ocal_3$ produces observable effects, corresponding to a right-handed quark current. Existing data
(from $\tau$ decays and $m_W$ measurements) implies $\Lambda \gesim 500\gev$ 

\begin{figure}
\vbox to 3 in{\epsfxsize=7 in\epsfbox[00 -530 612 262]{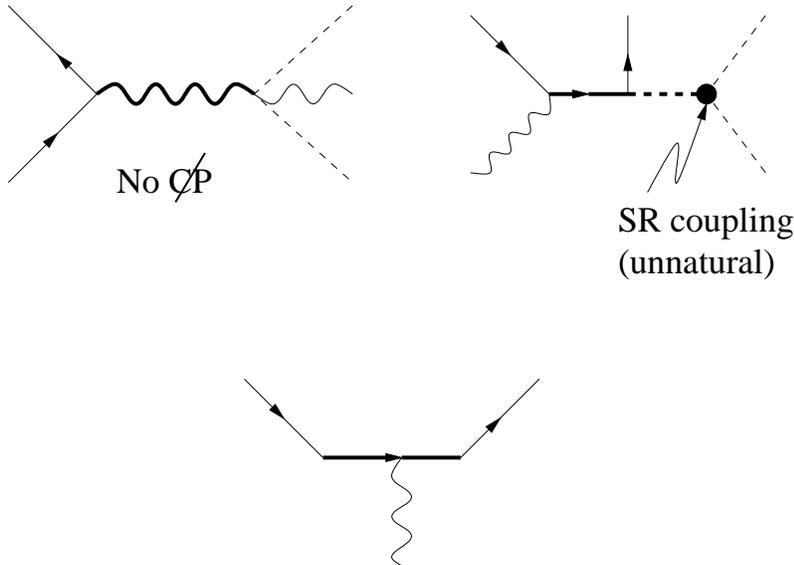}}
\label{f6}
\caption{Heavy physics contributing to CP violating operators. Wavy lines denote
vectors, solid lines fermions, and dashed ones scalars. Heavy lines denote
heavy excitations.}
\end{figure}

One can also determine the type of new interactions which might be probed using these
operators~\cite{aew2}. The heavy physics which can generate $ \ocal_3$ at tree level is
described in Fig. \ref{f6}. If the underlying theory is natural we conclude that
there will be no super-renormalizable couplings; in this case $\ocal_3$ will be generated
by heavy fermion exchanges only~\footnote{It is true that vertices involving light fermions,
light scalars and heavy fermions produce mixings between the light and heavy scales,
but this occurs at the one loop level. In contrast cubic terms of order $ \Lambda $
in the scalar potential would shift $v$ at tree level.}

Note finally that these arguments are only
valid for weakly coupled heavy physics. For strongly
coupled theories other CP violating operators can be important, e.g. 
\begin{equation}
 {\qwe\over\Lambda^2} B^{\mu\nu} \left(\bar e \gamma_\mu D_\nu e-\hbox{h.c} \right) 
\end{equation}
since $ |\qwe|\sim1$.

 \section{Other applications}

The \el~approach can be applied in many other situations such
as gravity and high temperature field theory. I will briefly consider the latter.

\subsection{Large temperatures}

It is a well-known fact that the thermodynamics of a system with Hamiltonian $H$
can be derived form the partition function $ \tr e^{-\beta H} $.
This resembles closely the (trace of the)
quantum evolution operator $ e^{-i H t } $ hence we can obtain the thermodynamics
of a system by the replacement
$ -i t \rightarrow \beta $: non-zero temperature field theory corresponds to
Euclidean field theory on a cylinder of perimeter$=\beta$, I will denote the
corresponding Euclidean time by $\tau$~\cite{kapusta}

Since the time direction is finite the fields are expanded in a Fourier series.
For bosons one obtains
\begin{equation}
\phi = \sum_{n=-\infty}^\infty \int{d^3k\over (2 \pi)^3}
\phi_n(\kk) e^{i(2 n \pi T \tau+\kk\cdot\rr)}
\end{equation}
and the corresponding free propagator is given by
\begin{equation}
\inv{(2 n \pi T)^2+\pp^2+m^2 }
\end{equation}
The field is periodic in $\tau$ due to the commutativity of the 
variables in the functional integral (there is a much more physical
reason, called the Kubo-Martin-Schwinger condition)~\cite{kapusta}.

Note that the $n \not=0$ modes become heavy as $T\to\infty$
so that in this limit only
the $n=0$ modes remain and the theory reduces to a
 3-D Euclidean field theory (there might be some subtleties involved, see below).

Fro fermions the expansion is in odd Fourier modes since the corresponding
integration variables anticommute. explicitly
\begin{equation}
\psi = \sum_{n=-\infty}^\infty \int{d^3k\over (2 \pi)^3}
\psi_n(\kk) e^{i(2 n+1)\pi T \tau+\kk\cdot\rr)}
\end{equation}
with fee propagator
\begin{equation}
 \inv{ \left[ i (2n+1) \pi T + \mu\right]\gamma^0 -\kk\cdot{\boldmath
\gamma} -m } 
\end{equation}
which shows that all modes become heavy as $T\to\infty$. There will
be then no fermions in the spectrum at very large temperatures. Note that this
occurs independently of the fermion mass~\cite{kapusta}.

Despite the absence of heavy fermions and scalars (effective mass $ \sim T$)
at large temperatures, we can still ask what is their effect on the scalar
modes that survive in this regime. To this end we can construct the corresponding
effective theory. I will illustrate the procedure using a simple example.

Consider the following scalar theory
\begin{equation}
 \lcal\up4 = 
\half \left(\partial\phi^2\right) -\half m^2 \phi^2 -{\lambda\over 4!}\phi^4 
\end{equation}
Then the excitations which survive at large $T$ are
\begin{equation}
\varphi(\xx)
= \sqrt{T} \; \int_0^\beta d\tau \; \phi(\xx,\tau)
\end{equation}
where $ \varphi$ is the dynamical variable of a 3 dimensional Euclidean field
theory (in 3 dimensions the scalar fields have units of $ \sqrt{\hbox{mass}}$
which explains the $ \sqrt{T}$ factor). The only symmetry (aside form Euclidean invariance)
is the reflection symmetry $\varphi \to - \varphi $. The scale of the new theory
is set by $\Lambda =T$, but in this case the model is supposed to describe physics
{\em above} $\Lambda $

With these considerations we can 
write the effective theory for $ \varphi $,
\begin{equation}
\leff = \half \left(\nabla \varphi \right)^2 + \half
a \varphi^2 +\inv{4!} b \varphi^4 +{c\over6!} \varphi^6 +O(1/T) 
\end{equation}
note that $b$ is a super-renormalizable coupling and may lead to infrared problems.

\begin{figure}
\vbox to 1.4in{\epsfxsize=6 in\epsfbox[-200 -660 412 132]{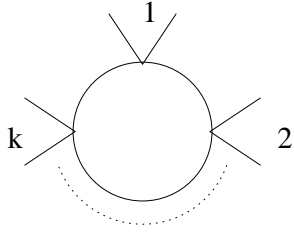}}
\label{f7}
\caption{Graphs exhibiting interesting infrared behavior at high temperatures}
\end{figure}

The coefficients $a$, $b$, $c$, etc. can be calculated
from the original theory. At one loop one obtains
\begin{eqnarray}
a={\lambda T^2\over24}\qquad
b=-{m\over2\pi} \left({\lambda T\over 4 m}\right)^2\qquad
c=\inv{4\pi} \left({\lambda T\over 4 m}\right)^3 \nonumber
\end{eqnarray}
But this calculation
has some potential problems. Consider the $2k$ point function 
at zero external
momentum; the corresponding graphs are given in Fig. \ref{f7}
A simple estimate (verified by explicit calculation) shows that
\begin{equation}
 \hbox{Graph} \propto 
\underbrace{{ \lambda^k \over m^{2k-4} } }_{\rm prefactors
+dim.~analysis} \quad\times\quad \underbrace{
\left( {T\over m}\right)^{k+1} }_{\rm integral+sums}
\end{equation}
which corresponds to the operator
\begin{equation}
 \ocal\up k \sim { \lambda^k \over m^{2k-4} } 
\left( {T\over m}\right)^{k+1}\left( \sqrt{T}\; \varphi\right)^{2k}\inv T
= m^3 \left( { \sqrt{\lambda}\; T \; \varphi \over m^{3/2} }\right)^{2k} 
\end{equation}
whose coefficient has {\em positive} powers of $\Lambda~(=T)$ and are not
suppressed at large temperatures. In fact, should this be correct the,
effective theory expansion would be useless.

\begin{figure}
\vbox to 1.4in{\epsfxsize=7 in\epsfbox[-50 -680 562 112]{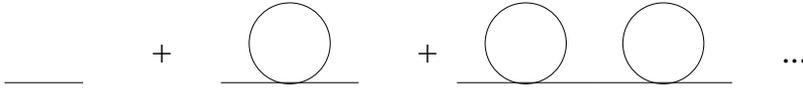}}
\label{f8}
\caption{Radiative corrections to the $n=0$ propagator
which cure the infrared divergences in the effective coefficients.}
\end{figure}

The solution to this infrared problem (diverging effective coefficients as
$m \to 0 $ is well known for this type of theories~\cite{kapusta}: the
propagator for the $n=0$ mode gets dressed and in so doing the $m^2$
gets shifted by an amount $ \propto T^2$. Explicitly,
the graphs in Fig. \ref{f8} shift 
\begin{equation}
 m^2 \to m^2 +{ \lambda\over24}T^2
\end{equation}
so that the previous expression for the effective
operator coefficient becomes
\begin{equation}
 \ocal\up k \sim {\lambda^{(3-k)/2} \over T^k} \varphi^{2k} 
\end{equation}
which vanishes as $ T \to \infty $. Note that there is still
a remnant of the infrared properties of the theory in that
the coefficients still diverge as $ \lambda \to 0 $.

\paragraph{QCD at high temperatures}

The previous arguments can be applied to the case of gauge theories.
Just as for the scalar field, the gauge field is periodic in $\beta $
and can be expanded in Fourier modes. At high temperatures, all but 
the $n=0$ modes are heavy with masses $ \sim T $. The remaining light
modes are
\begin{equation} 
\AA^A_{n=0}\equiv \aa^A \qquad A^0{}^A_{n=0}\equiv \varphi^A
\end{equation}
leaving a 3-D Euclidean $\su3$ model with gauge fields
$ \aa^A $ and with a scalar octet (the $ \varphi^A$).
The 3-D gauge coupling constant is $ g\sqrt{T} $ (where
$g$ denotes the QCD gauge coupling)

The simplest infrared divergences are cured by the 
dressing the gluon propagator at one loop~\cite{kapusta}; 
the $ \phi^A$ propagator at large $T$ then becomes
\begin{equation}
\inv{p^2} \to \inv{p^2+c g^2 T^2}
\end{equation}
for some numerical constant $c$. 
But this effect is not
extended to the $\aa^A$ for the corresponding vacuum
polarization obeys $\Pi_{ii}(p\to0)\to0$~\cite{kapusta}.

\begin{figure}
\vbox to 1.5in{\epsfxsize=8 in\epsfbox[-100 -680 512 112]{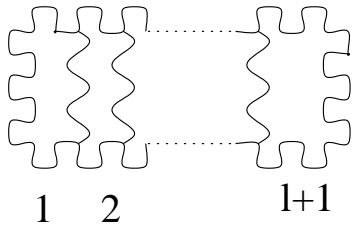}}
\label{f9}
\caption{Some radiative corrections to the QCD free energy}
\end{figure}

The fact that the \aa\ remain massless leads to various
interesting  problems. For example 
the higher order corrections to the free energy,
provided by graphs in Fig. \ref{f9}. Suppose that the
gauge bosons have a (dynamically generated) mass $m$. In this case
a graph with $\ell$ loops behaves as~\cite{kapusta}
\begin{equation}
g^6 T^4 (g^2 T/m)^{\ell-3}\qquad (\ell>3)
\end{equation}

For the case where internal lines correspond to
$A^0$ (or $\varphi^A$) $ m \sim g T $ and the graph
is well behaved, $ \sim g^{ \ell+3 } T^4 $. On the other
hand when the internal lines represent $ A^i $ (or, equivalently,
$\aa^A$) propagator a problem will arise unless 
$ m \sim g^2 T $ is generated (we already know there is
no $O(g)$ correction to $m$). This so-called magnetic mass
has not been obtained perturbatively though it is widely believed
to be generated.

Additional problems arise since the gauge coupling constant
in the 3-d theory has dimensions of $ \sqrt{\hbox{mass}}$
leading to super-renormalizable interactions with the
related infrared divergences~\cite{braaten}.

\section{Conclusions}

In these lectures I have provided a review of some of the very many
aspects and properties of effective theories, as well as some of their
application. Despite this drawback I hope it does give a flair for the
strength of the approach.

Effective theories  will be used in deriving the implications of new
data on the properties of the physics which underlies the \sm, but in
addition it can be applied to a wide variety of phenomena ranging form 
QCD to superconductivity. It is this flexibility which makes the
formalism so attractive.

\end{document}